\documentstyle[prd,aps]{revtex}

\newcommand{\beq}{\begin{equation}}
\newcommand{\eeq}{\end{equation}}
\newcommand{\bea}{\begin{eqnarray}}
\newcommand{\eea}{\end{eqnarray}}
\newcommand{\bmath}{\begin{mathletters}}
\newcommand{\emath}{\end{mathletters}}

\newcommand{\Tr}{\hbox{Tr}}
\renewcommand{\Re}{ {\rm Re}\, }
\renewcommand{\Im}{ {\rm Im}\, }

\newcommand{\SF}[2]{ \overline{(#1 \otimes #2)} }
\newcommand{\Slash}[1]{ \!\not{\!\!#1} }
\newcommand{\Back}[1]{\stackrel{\scriptstyle{\leftarrow}}{#1}}
\newcommand{\Bk}[1]{\stackrel{\scriptstyle{\leftarrow}}{#1}\!{}^2}

\newcommand{\basispl}{
   \put(-.5,-.5){\line(1,0){1}}
   \put(.5,-.5){\line(0,1){1}}
   \put(.5,.5){\line(-1,0){1}}
   \put(-.5,.5){\line(0,-1){1}}}

\newcommand{\Plaquette}
   {\setlength{\unitlength}{.5cm}\raisebox{-.2cm}{
   	\begin{picture}(1.2,1.2)(-.6,-.6)
   		\basispl
   		\put(-.5,-.5){\circle*{.2}}
   		\put(-.5,.5){\circle*{.2}}
   		\put(.5,-.5){\circle*{.2}}
   		\put(.5,.5){\circle*{.2}}
   	\end{picture}}
   }
\newcommand{\LONGpl}
   {\setlength{\unitlength}{.5cm} \raisebox{-.2cm}{
   	\begin{picture}(2.2,1.2)(-1.1,-.6)
   		\put(-1,-.5){\line(1,0){2}}
   		\put(-1,.5){\line(1,0){2}}
   		\put(-1,-.5){\line(0,1){1}}
   		\put(1,-.5){\line(0,1){1}}
   		\multiput(-1,-.5)(1,0){3}{\circle*{.2}}
   		\multiput(-1,.5)(1,0){3}{\circle*{.2}}
        \end{picture}}
   }
\newcommand{\CornerPl}
   {\setlength{\unitlength}{.5cm} \raisebox{-.3268cm}{
   	\begin{picture}(1.7071,1.7071)(-.7071,-.7071)
   		\put(-.7071,-.7071){\line(0,1){1}}
   		\put(0,1){\line(1,0){1}}
   		\put(1,1){\line(0,-1){1}}
   		\put(-.7071,-.7071){\line(1,0){1}}
   		\put(0,1){\line(-1,-1){.7071}}
   		\put(1,0){\line(-1,-1){.7071}}
   		\put(-.7071,-.7071){\circle*{.1}}
   		\put(-.7071,.2929){\circle*{.2}}
   		\multiput(0,0)(1,0){2}{\circle*{.2}}
   		\multiput(0,1)(1,0){2}{\circle*{.2}}
   		\multiput(-.7071,-.7071)(1,0){2}{\circle*{.2}}
   		\multiput(0,0)(.25,0){4}{\circle*{.03}}
   		\multiput(0,0)(0,.25){4}{\circle*{.03}}
   		\multiput(0,0)(-.1768,-.1768){4}{\circle*{.03}}
   	\end{picture}}
   }
\newcommand{\FoldedPl}
   {\setlength{\unitlength}{.5cm} \raisebox{-.3268cm}{
   	\begin{picture}(1.7071,1.7071)(-.7071,-.7071)
   		\put(0,0){\line(0,1){1}}
   		\put(0,1){\line(1,0){1}}
   		\put(1,1){\line(0,-1){1}}
   		\put(-.7071,-.7071){\line(1,0){1}}
   		\put(0,0){\line(-1,-1){.7071}}
   		\put(1,0){\line(-1,-1){.7071}}
   		\multiput(0,0)(1,0){2}{\circle*{.2}}
   		\multiput(0,1)(1,0){2}{\circle*{.2}}
   		\multiput(-.7071,-.7071)(1,0){2}{\circle*{.2}}
   		\multiput(0,0)(.25,0){4}{\circle*{.03}}
   	\end{picture}}
   }

\begin{document}
\draft
\input epsf

\begin{titlepage}
Feb. 1997       \hfill  CU-TP-816
\vskip 1.2 in
\begin{center}
{\Large\bf On-shell improved lattice QCD with staggered fermions}
\\\vspace{2cm}
{\bf Yubing Luo} \\\vspace{1cm}
        Columbia University, Department of Physics, \\
	New York, NY 10027 \\
E - mail: roy@cuphyc.phys.columbia.edu \\
\end{center}
\vspace{2cm}


\begin{abstract} 

By using Symanzik's improvement program, we study on-shell
improved lattice QCD with staggered fermions. We find that
there are as many as 15 independent lattice operators of dimension
of six~(including both gauge and fermion operators) which must be
added to the unimproved action to form an $O(a^2)$ improved action. 
Among them, the total number of
dimension-6 gauge operators and fermion bilinears is 5.
The other 10 terms are four-fermion operators.
At the tree-level and tadpole-improved tree-level,
all the 10 four fermion operators are absent.

\bigskip
\bigskip
\end{abstract}


\end{titlepage}

\section{Introduction} {\label{sect:1}}

In recent years, there has been a surge in developing
and applying improved actions for the numerical simulations
of lattice QCD. Up to now, most research has focused on
improvement of Wilson fermions in an effort to reduce the
$O(a)$ cutoff effects in the simulations. On the other hand,
the absence of $O(a)$ errors for the staggered fermion
action~\cite{Sharpe:1,Luo:1} and the
complexity of the staggered formalism mean that its $O(a^2)$
improvement has received little attention.
Almost ten years ago, Naik proposed adding a third-nearest-neighbor
term to the standard staggered fermion action to remove some
$O(a^2)$ effects~\cite{Naik}. His study was based on the
Dirac-K\"ahler equation, not on the standard staggered formulation.
Although these two fermion formulations are the same in the free case,
they are quite different when the gauge interactions are included
and the difference is of the order of $a^2$.
So, the Naik term may not remove all $O(a^2)$ errors from the
simulations by using staggered fermions.
This statement was demonstrated by the recent
numerical simulation from the MILC 
group~\cite{MILC:1,MILC:2}

One approach to improvement is to construct a perfect
action~\cite{Perfect_action:1}. The classical perfect action
for free staggered fermions was already proposed in
ref.~\cite{Perfect_staggered:1}.
In this paper,
we will apply Symanzik's improvement scheme to staggered
fermions and discuss its on-shell improvement through $O(a^2)$.
We will show that including only the Naik term in the improved
staggered fermion action is not enough to remove all
order $a^2$ errors from on-shell quantities. 
Meanwhile, because both the standard
Wilson gauge action and the standard staggered fermion action have
$O(a^2)$ errors, we must improve both of them at the same time. We
will show that these two improvements are not independent but
connected by an isospectral transformation of the gauge fields.
The recent calculations of the MILC and Bielefeld~\cite{Bielefeld:1}
groups can be easily explained by the result of our analysis.


This paper is organized as follows.
In section \ref{sect:2},
we will discuss the $O(a^2)$ improvement of the
staggered fermion action by finding all linearly independent
dimension-6 operators following Symanzik's scheme.
In section \ref{sect:3}, we will give the coefficients in tree level
up to order $a^2$.
In section \ref{sect:4}, we will give the general isospectral
transformation of fermion fields and gauge fields and give the
form of the simplified on-shell improved action.  Section \ref{sect:5}
is the conclusion.
The computations of the tree-level coefficients are presented by two
appendices.

\section{Possible counterterms} {\label{sect:2}}

When Symanzik's improvement scheme is applied to construct
an $O(a^2)$ improved lattice action,
the first step is to find all dimension
six operators which are scalars under the lattice symmetry
group. These operators, treated as counterterms, are then added to
the action to remove all
$O(a^2)$ errors from physical quantities.
Before doing that, we will introduce some notation which will
simplify our presentation. For
the transformation properties of staggered fermion fields,
the reader is requested to consult ref.~\cite{Luo:1}, which we
will refer to as paper\,(I) in the following.

\subsection{Definitions and Notations}

The whole lattice can be viewed as being composed of elementary
hypercubes consisting of 16 lattice sites. We will use $x$ to label
the individual lattice sites and $y$, which
has only even coordinates,
to label each hypercube. A site inside a hypercube is
represented by a ``hypercubic vector'' $A$, whose components
can only take the values of either $0$ or $1$. The relationship
between these 3 vectors is given by
\beq
    x = y + A.
\eeq

The hypercubic fields are defined as:
\bmath
\beq
    \chi_A(y) = \frac{1}{4}\chi(y+A),
\eeq
\beq
    \bar{\chi}_A(y) = \frac{1}{4}\bar{\chi}(y+A).
\eeq
With the notation
\beq
    \bar{\chi}\,{\cal{M(U)}}\,\chi
    = \sum_{y,AB} \bar{\chi}_A(y)\,{\cal{M(U)}}_{AB}\,\chi_B(y),
\eeq
and
\beq
    \overline{(\gamma_S \otimes \xi_F)}_{AB}
    = \frac{1}{4}\Tr(\gamma_A^\dagger \gamma_S \gamma_B
    	\gamma_F^\dagger),
\eeq
we can write the standard staggered fermion action in a compact
form as
\beq
    S_F = \bar{\chi} \left[ \sum_\mu \SF{\gamma_\mu}{I}
    	{\cal{D}}_\mu + m \right] \chi .
\eeq
\emath
Furthermore, when discussing the fermion operators, 
we will use the following notation:
\bmath
\beq
    \Slash{\cal{D}} = \sum_\mu \SF{\gamma_\mu}{I}{\cal{D}}_\mu,
\eeq
\beq
    {\cal{D}}^2 = \sum_\mu \SF{\gamma_\mu}{I}{\cal{D}}_\mu
    	\,\SF{\gamma_\mu}{I}{\cal{D}}_\mu,
\eeq
\beq
    {\cal{D}}^3 = \sum_\mu \SF{\gamma_\mu}{I}{\cal{D}}_\mu
    	\,\SF{\gamma_\mu}{I}{\cal{D}}_\mu
    	\,\SF{\gamma_\mu}{I}{\cal{D}}_\mu,
\eeq    
\beq
    {\cal{D}}\Slash{\cal{D}}{\cal{D}} = \sum_{\mu\nu}
    	\SF{\gamma_\mu}{I}{\cal{D}}_\mu
    	\,\SF{\gamma_\nu}{I}{\cal{D}}_\nu
    	\,\SF{\gamma_\mu}{I}{\cal{D}}_\mu.
\eeq
\emath

\subsection{Fermion bilinears}

The lattice symmetry group of staggered fermion
action~\cite{symmetry:1,Luo:1} includes:
translation, reflection, rotation, charge conjugate and a continuous
$U_V(1)$. When the mass parameter $m$ is 0, there is a second continuous
$U_A(1)$ symmetry.

When $m = 0$, we can identify the following five
independent operators which are scalars under all symmetry
transformations including $U_A(1)$:
\bmath
\beq
    {\cal{O}}_1 = \bar{\chi}\,{\cal{D}}^3 \chi,
\eeq
\beq
    {\cal{O}}_2 = \bar{\chi}\,\frac{1}{2} \left(
    	{\cal{D}}^2\Slash{\cal{D}} -\Slash{\cal{D}}{\cal{D}}^2
	\right)\chi,
\eeq
\beq
    {\cal{O}}_3 = \bar{\chi}\, \frac{1}{2} \left(
    	{\cal{D}}^2\Slash{\cal{D}} + \Slash{\cal{D}}{\cal{D}}^2
	- 2 \Slash{\cal{D}}^3 \right)\chi,
\eeq
\beq
    {\cal{O}}_4 = \bar{\chi}\, \left(
    	{\cal{D}}^2\Slash{\cal{D}} + \Slash{\cal{D}}{\cal{D}}^2
	- 2 {\cal{D}}\Slash{\cal{D}}{\cal{D}} \right)\chi,
\eeq
\beq
    {\cal{O}}_5 = \bar{\chi}\,\Slash{\cal{D}}^3 \chi.
\eeq
\emath
For the case of non-zero fermion mass, the $U_A(1)$ symmetry is
violated, and there are two more allowed counterterms:
\bmath
\beq
    {\cal{O}}_6 = m\,\bar{\chi}\,\Slash{\cal{D}}^2 \chi,
\eeq
\beq
    {\cal{O}}_7 = m\,\bar{\chi}\,{\cal{D}}^2 \chi.
\eeq
\emath

\subsection{Four fermion operators}

When considering $O(a^2)$ corrections to the fermion action, we must
examine not only dimension-6 operators bilinear in the fermion
fields, but also four fermion operators of dimension of six.

Using hypercubic coordinates, we can connect the spin and flavor
indices in staggered four fermion operators in combinations of the
form:
\beq
    \bigl(\bar\chi \,{\cal{M(U)}}\,\chi\bigr)^2
    = \sum_y\,\sum_{AB} \bar{\chi}_A(y) {\cal{M(U)}}_{AB} \chi_B(y)\,
    	\sum_{CD} \bar{\chi}_C(y){\cal{M(U)}}_{CD}\chi_D(y).
\eeq
However, the color indices in such an operator might be combined in
four ways:
\bmath
\beq {\label{color_form:1}}
    \bar\chi_a \delta_{aa'} \Bigl({\cal{M(U)}}\chi\Bigr)_{a'}\,
    \bar\chi_b \delta_{bb'} \Bigl({\cal{M(U)}}\chi\Bigr)_{b'},
\eeq
\beq {\label{color_form:2}}
    \bar\chi_a \delta_{ab'} \Bigl({\cal{M(U)}}\chi\Bigr)_{a'}\,
    \bar\chi_b \delta_{ba'} \Bigl({\cal{M(U)}}\chi\Bigr)_{b'},
\eeq
\beq {\label{color_form:3}}
    \bar\chi_a t^i_{aa'} \Bigl({\cal{M(U)}}\chi\Bigr)_{a'}\,
    \bar\chi_b t^i_{bb'} \Bigl({\cal{M(U)}}\chi\Bigr)_{b'},
\eeq
\beq {\label{color_form:4}}
    \bar\chi_a t^i_{ab'} \Bigl({\cal{M(U)}}\chi\Bigr)_{a'}\,
    \bar\chi_b t^i_{ba'} \Bigl({\cal{M(U)}}\chi\Bigr)_{b'},
\eeq
\emath
with
\beq
    t^i = \frac{\lambda^i}{2},
\eeq
where $\lambda^i$ are the $SU(3)$ Gell-Mann matrices, and
as usual, the repetition of the indices $a,a',b,b'$
and $i$ means summation. Because of the completeness relation of
the matrices $\lambda^i$
\beq
    \sum_{i=1}^{8} \lambda^i_{aa'} \lambda^i_{bb'}
    = 2 \delta_{ab'}\delta_{ba'}
      - \frac{2}{3}\delta_{aa'}\delta_{bb'},
\eeq
the operators with the form of Eq.~(\ref{color_form:1}) and
Eq.~(\ref{color_form:2}) can be expressed as linear
combinations of operators
with the form of Eq.~(\ref{color_form:3}) and
Eq.~(\ref{color_form:4}).
Furthermore, the operators with the form of Eq.~(\ref{color_form:4})
can be expressed in terms of the operators with the form of
Eq.~(\ref{color_form:3}) by making a Fierz transformation.
Hence, we need only consider the operators with the form of
Eq.~(\ref{color_form:3}).\footnote{If additional, explicit flavors
of staggered fermions are introduced, Fierz symmetry cannot be used
a second time, so we will need to introduce both flavor
adjoint and singlet fermion bilinears, effectively doubling
the number of flavor singlet,
four fermion operators that must be considered.}

For convenience, we will not write out the links explicitly in
the remaining part of this
section unless there would otherwise be confusion.
After applying the staggered fermion symmetry transformation
including rotation, reflection, charge conjugate and
the continuous $U_V(1)\times U_A(1)$, we found there are
18 operators which are invariant:
\bmath	{\label{four_op:hypercubic}}
\beq
    {\cal F}_1 = 
    \Bigl(\bar\chi t^a\SF{I}{I}\chi\Bigr)^2 -
    	\Bigl(\bar\chi t^a\SF{\gamma_5}{\xi_5}\chi\Bigr)^2 +
    \sum_\mu\Bigl[\Bigl(\bar\chi t^a
    	\SF{\gamma_\mu}{\xi_\mu}\chi\Bigr)^2 -
    	\Bigl(\bar\chi t^a\SF{\gamma_{5\mu}}
	{\xi_{5\mu}}\chi\Bigr)^2\Bigr],
\eeq
\beq
    {\cal F}_2 = 
    \Bigl(\bar\chi t^a\SF{I}{I}\chi\Bigr)^2 -
    	\Bigl(\bar\chi t^a\SF{\gamma_5}{\xi_5}\chi\Bigr)^2 -
    \sum_\mu\Bigl[\Bigl(\bar\chi t^a
    	\SF{\gamma_\mu}{\xi_\mu}\chi\Bigr)^2 -
    	\Bigl(\bar\chi t^a\SF{\gamma_{5\mu}}
	{\xi_{5\mu}}\chi\Bigr)^2\Bigr],
\eeq
\beq
    {\cal F}_3 = 
    \sum_\mu\Bigl(\bar\chi t^a\SF{\gamma_{5\mu}}{I}\chi\Bigr)^2,
\eeq
\beq
    {\cal F}_4 = 
    \sum_\mu\Bigl(\bar\chi t^a\SF{\gamma_\mu}{\xi_5}\chi\Bigr)^2,
\eeq
\beq
    {\cal F}_5 = 
    \sum_{\mu\neq\nu\neq\lambda} \Bigl(\bar\chi
    	 t^a\SF{\gamma_{\mu\nu}}{\xi_\lambda}\chi\Bigr)^2,
\eeq
\beq
    {\cal F}_6 = 
    \sum_{\mu\neq\nu} \Bigl(\bar\chi
    	 t^a\SF{\gamma_{\mu\nu}}{\xi_{5\nu}}\chi\Bigr)^2,
\eeq
\beq
    {\cal F}_7 = 
    \sum_{\mu\neq\nu\neq\lambda} \Bigl(\bar\chi
    	 t^a\SF{\gamma_\mu}{\xi_{\nu\lambda}}\chi\Bigr)^2,
\eeq
\beq
    {\cal F}_8 = 
    \sum_{\mu\neq\nu} \Bigl(\bar\chi
    	 t^a\SF{\gamma_{5\mu}}{\xi_{\mu\nu}}\chi\Bigr)^2,
\eeq
\beq
    {\cal F}_9 = 
    \sum_\mu\Bigl(\bar\chi t^a\SF{I}{\xi_{5\mu}}\chi\Bigr)^2,
\eeq
\beq
    {\cal F}_{10} = 
    \sum_\mu\Bigl(\bar\chi t^a\SF{\gamma_5}{\xi_\mu}\chi\Bigr)^2,
\eeq
\beq
    {\cal F}_{11} = 
    \sum_\mu\Bigl(\bar\chi t^a\SF{\gamma_\mu}{I}\chi\Bigr)^2,
\eeq
\beq
    {\cal F}_{12} = 
    \sum_{\mu\neq\nu\neq\lambda} \Bigl(\bar\chi
    	 t^a\SF{\gamma_{\mu\nu}}{\xi_{5\lambda}}\chi\Bigr)^2,
\eeq
\beq
    {\cal F}_{13} = 
    \sum_\mu\Bigl(\bar\chi t^a\SF{I}{\xi_\mu}\chi\Bigr)^2,
\eeq
\beq
    {\cal F}_{14} = 
    \sum_{\mu\neq\nu\neq\lambda} \Bigl(\bar\chi
    	 t^a\SF{\gamma_{5\mu}}{\xi_{\nu\lambda}}\chi\Bigr)^2,
\eeq
\beq
    {\cal F}_{15} = 
    \sum_\mu\Bigl(\bar\chi t^a
    	\SF{\gamma_{5\mu}}{\xi_5}\chi\Bigr)^2,
\eeq
\beq
    {\cal F}_{16} = 
    \sum_{\mu\neq\nu} \Bigl(\bar\chi
    	 t^a\SF{\gamma_{\mu\nu}}{\xi_\nu}\chi\Bigr)^2,
\eeq
\beq
    {\cal F}_{17} = 
    \sum_\mu\Bigl(\bar\chi t^a
    	\SF{\gamma_5}{\xi_{5\mu}}\chi\Bigr)^2,
\eeq
\beq
    {\cal F}_{18} = 
    \sum_{\mu\neq\nu} \Bigl(\bar\chi
    	 t^a\SF{\gamma_\mu}{\xi_{\mu\nu}}\chi\Bigr)^2.
\eeq
\emath
These operators are not invariant under translation. However,
the additional terms generated by translations
are of $O(a)$. For any operator listed above,
we can combine it with some higher dimension operator so
that the new operator is invariant under translation. 
This new operator differs with the old one by an $O(a)$ term and
both of them have the same continuum form.
Therefore the translation symmetry does not reduce the number of
invariant operators here.

After adding some higher dimensional terms,
we can make the 18 four fermion operators
listed above invariant under translation and rewrite
them in terms of the fields $\chi(x)$ and $\bar\chi(x)$.
First,
\beq	{\label{chi_form:1}}
    {\cal{F}}_1 = \sum_{x,a} \bar\chi(x) t^a\chi(x)
    	\sum_e \bar\chi(x\!+\!e)t^a\chi(x\!+\!e),
\eeq
where the sum over $e$ is a sum over the 8 possible lattice
displacements of length ``1''.
Second,
\beq	{\label{chi_form:2}}
    {\cal{F}}_2 = \sum_{x,a} \bar\chi(x) t^a\chi(x)
    	\sum_v \bar\chi(x\!+\!v)t^a\chi(x\!+\!v),
\eeq
where the sum over $v$ is over the 32 possible lattice displacements
of length ``$\sqrt{3}$~''.
Next,
\bea	{\label{chi_form:3}}
    {\cal{F}}_i = \sum_{x,a} \sum_\mu
    	{\cal C}^a_\mu(x) \frac{1}{256}
	\sum_c w(c)\eta_5(c) P^{(i)}_\mu(c) {\cal C}^a_\mu(x+c)
		\\
    i = 3, \cdots , 10.		\nonumber
\eea
This equation contains a number of new elements which we will now
define. The sum over $c$ is a sum over the 81 displacements with
coordinates $c_\mu = -1, 0, 1$. The weight is:
\beq
    w(c) = \prod_{\mu=1}^{4}\,(2-\left|c_\mu\right|).
\eeq
The fermion bilinear ${\cal C}^a_\mu(x)$ is given by:
\beq
    {\cal C}^a_\mu(x) = \bar\chi(x) t^a \sum_{v\perp\hat\mu}
	\chi(x\!+\!v),
\eeq
where the sum is over the 8 possible lattice displacements of length
``$\sqrt{3}$~'' which are perpendicular to $\hat\mu$ direction.
The phase factors $P^{(i)}_\mu(c)$ are defined by
\bea
    P^{(3)}_\mu(c) = \eta_\mu(c), \qquad \qquad \qquad
    P^{(4)}_\mu(c) = \varepsilon(c)\eta_\mu(c),
		\nonumber \\
    P^{(5)}_\mu(c) = \varepsilon(c)\tau_\mu(c)\eta_\mu(c),  \qquad
    P^{(6)}_\mu(c) = \tau_\mu(c)\eta_\mu(c),
		\nonumber \\
    P^{(7)}_\mu(c) = \tau_\mu(c)\zeta_\mu(c), \qquad
    P^{(8)}_\mu(c) = \varepsilon(c)\tau_\mu(c)\zeta_\mu(c),
		\nonumber \\
    P^{(9)}_\mu(c) = \varepsilon(c)\zeta_\mu(c), \qquad \qquad \qquad
    P^{(10)}_\mu(c) = \zeta_\mu(c),
\eea
where
\bea
    \tau_\mu(c) = \frac{1}{3}\sum_{\nu\neq\mu}(-1)^{c_\nu},\nonumber \\
    \eta_\mu(c) = (-1)^{c_1 + \cdots + c_{\mu-1}},\nonumber \\
    \zeta_\mu(c) = (-1)^{c_{\mu+1}+\cdots + c_4},\nonumber \\
    \varepsilon(c) = (-1)^{c_1 +\cdots + c_4},\nonumber	\\
    \eta_5(c) = \prod_{\mu=1}^{4}\,\eta_\mu(c).
\eea
The remaining 10 operators can be written as:
\bea	{\label{chi_form:4}}
    {\cal{F}}_i = \sum_{x,a} \sum_{\mu} {\cal B}^a_\mu(x)
	\frac{1}{256}\sum_c w(c)P^{(i)}_{\mu}(c){\cal B}^a_\mu(x+c),
		\\
    i = 11, \cdots , 18,		\nonumber
\eea
with the fermion bilinear
\beq
    {\cal B}^a_\mu(x) = \frac{1}{2}\Bigl[
        \bar\chi(x) t^a\chi(x\!+\!\hat\mu)
        + \bar\chi(x) t^a\chi(x\!-\!\hat\mu) \Bigr]
\eeq
and the phase factors are given by
\bea
    P^{(11)}_{\mu}(c) = \eta_\mu(c),
    	~~~\qquad \qquad \qquad
    P^{(12)}_{\mu}(c) = \tau_\mu(c)\eta_\mu(c),
    	\nonumber \\
    P^{(13)}_{\mu}(c) = \zeta_\mu(c),
    	~~~\qquad \qquad \qquad
    P^{(14)}_{\mu}(c) = \tau_\mu(c)\zeta_\mu(c),
	\nonumber \\
    P^{(15)}_{\mu}(c) = \varepsilon(c)\eta_\mu(c),
    	\qquad \qquad
    P^{(16)}_{\mu}(c) = \varepsilon(c)\tau_\mu(c)\eta_\mu(c),
	\nonumber\\
    P^{(17)}_{\mu}(c) = \varepsilon(c)\zeta_\mu(c),
    	\qquad \qquad
    P^{(18)}_{\mu}(c) = \varepsilon(c)\tau_\mu(c)\zeta_\mu(c),
    	\nonumber
\eea

We have now discussed all dimension-6 fermion operators
which are invariant under the lattice symmetry group.
Therefore, we can write down a suitable
$O(a^2)$ improved staggered fermion action as:
\beq {\label{improved_fermion_action}}
    S_F = \bar{\chi}\, \left(\,\Slash{\cal{D}}+m \right)\,\chi
    	+ a^2\,\sum_{i=1}^{7} b_i(g_0^2, ma){\cal{O}}_i
	+ a^2\,\sum_{i=1}^{18}b'_i(g_0^2, ma){\cal{F}}_i.
\eeq

The reality of the action requires that $b_2$ is imaginary, and all
other $b'$ and $b$'s are real.

\subsection{Gauge fields}

The Symanzik improvement of the gauge theory action
was studied more than a decade
ago~\cite{Weisz:1,Luscher:2}.
It was found that there are three independent six-link
products which must be added to the
original Wilson action to form an $O(a^2)$ improved gauge action.
The improved gauge action can be written as:
\beq{\label{improved_gauge_action}}
    S_G[U] = \sum_{i=0}^3 c_i(g_0^2) {\cal L}_i
\eeq
where the link product ${\cal L}_i$ are defined as:
\bmath
\beq
    {\cal L}_0 = \frac{\beta}{3}\sum_x~\Re\Tr\Bigl<1 -
    	\Plaquette~\Bigr>,
\eeq
\beq
    {\cal L}_1 = \frac{\beta}{3}\sum_x~\Re\Tr\Bigl<1 -
    	\LONGpl~\Bigr>,
\eeq
\beq
    {\cal L}_2 = \frac{\beta}{3}\sum_x~\Re\Tr\Bigl<1 -
    	\CornerPl~\Bigr>,
\eeq
\beq
    {\cal L}_3 = \frac{\beta}{3}\sum_x~\Re\Tr\Bigl<1 -
    	\FoldedPl~\Bigr>,
\eeq
\emath
where $\langle ~\rangle$ implies an average over orientations.
The four $c_i$'s satisfy the normalization condition:
\beq
    c_0(g_0^2) + 8 c_1(g_0^2) + 8 c_2(g_0^2)
    	+ 16 c_3(g_0^2) = 1.
\eeq

For on-shell improved pure gauge theory, it was shown that we can set
$c_3(g_0^2)$ to zero by a change of field variable in the path
integral. However, we have to be careful when we discuss an
improved action which includes the quarks, because the change of
gauge field variable will also have an impact on the fermion action.
We will discuss this issue in the latter part of this paper when we
discuss the isospectral transformations.

\section{Tree level improvement}{\label{sect:3}}

A natural way to do the tree level improvement is to expand the
lattice action to order $a^2$ and to adjust the coefficients
$b_i$ so that the difference from the continuum Lagrangian
is of order of $a^3$. This also improves the free fermion
propagator through order of $a^2$.

Define the gauge covariant hypercubic fermion fields as
\bmath \label{cov_field}
\beq
    \varphi_A(y)={\cal{U}}_A(y)\,\chi_A(y),
\eeq
\beq
    {\bar{\varphi}}_A(y)={\bar{\chi}}_A(y){\cal{U}}^\dagger_A(y),
\eeq
\emath
where ${\cal{U}}_A(y)$ is the average of link products along
the shortest paths from $y$ to $y+A$.
For the classical continuum limit of the standard staggered
fermion action~(see Appendix \ref{Append:A}), we find:
\bea \label{classical_limit}
    S_F = \int_y \sum_{AB}~{\bar{\varphi}}_A(y) \biggl\{
    	\sum_\mu \overline{(\gamma_\mu \otimes I)}_{AB}D_\mu +
	m~\overline{(I \otimes I)}_{AB}	\nonumber       \\
      + a\sum_\mu\Bigl[ig_0
	\sum_\lambda A_\lambda \overline{(\gamma_\mu
	\otimes I)}_{AB} F_{\lambda\mu}
	- \overline{(\gamma_5 \otimes \xi_{5\mu})}_{AB} D_\mu^2
	\Bigr] \nonumber	\\
      + \frac{2a^2}{3}\sum_\mu\overline{(\gamma_\mu
        \otimes I)}_{AB} D_\mu^3 \nonumber	\\
      + \frac{i}{2}g_0 a^2 \Bigl[\sum_{\mu\nu\lambda} A_\lambda A_\nu
        \overline{(\gamma_\mu \otimes I)}_{AB}\left[D_\nu,
	F_{\lambda\mu}\right]	\nonumber       \\
      - \sum_{\mu\lambda} A_\lambda
	\overline{(\gamma_5 \otimes \xi_{5\mu})}_{AB} \bigl(
	\left[D_\mu, F_{\lambda\mu}\right] + 3F_{\lambda\mu}D_\mu
	\bigr) \Bigr] \biggr\} {\varphi}_B(y)
	\nonumber	\\
      + O(a^3),
\eea
where $D_\mu$ is the continuum covariant derivative, and $F_{\mu\nu}$
is the continuum field strength.

Now, let us define a new set of fermion field variables
\bmath
\label{new_field}
\beq
    \phi_A=\exp\Bigl(-a\sum_\lambda A_\lambda \overline{D}_\lambda
    	\Bigr)~\varphi_A,
\eeq
\beq
    {\bar{\phi}}_A={\bar{\varphi}}_A~\exp\Bigl(-a\sum_\lambda 
    	A_\lambda \Back{\overline{D}}_\lambda\Bigr),
\eeq
\emath
where $\overline{D}$ is defined in Eq.~(\ref{cov_derivative:D}).
This definition is an obvious extension to higher order in $a$ of
the order $a$ transformation discussed in paper\,(I)
chosen to remove extraneous terms of $O(a)$ which appear when the
original staggered fermion action is written in terms of hypercubic
variables.
The continuum limit of the staggered fermion action in terms of the
$\phi$ field can be written as
\beq
    S_F = \int_y \sum_{AB}~\bar{\phi}_A(y) \Bigl\{ \sum_\mu
        \overline{(\gamma_\mu \otimes I)}_{AB}
	\bigl[ D_\mu + \frac{a^2}{6}D_\mu^3 \bigr]
	+ m~\overline{(I \otimes I)}_{AB} \Bigr\}
	\phi_B(y) + O(a^3).
\eeq
From this equation, it is easy to get the tree-level values of the
coefficients occurring in Eq.~(\ref{improved_fermion_action}).
We obtain:
\beq
    b_1(0, ma) = -\frac{1}{6},	\qquad
\eeq
and all other $b$ and $b'$ are zero.

The coefficients in Eq.~(\ref{improved_gauge_action}) were given
in references~\cite{Weisz:1} and~\cite{Luscher:2}.
Their values are:
\beq
    c_0(0) = \frac{5}{3}, \qquad 
    c_1(0) = -\frac{1}{12}, \qquad
    c_2(0) = c_3(0) = 0.
\eeq

The tree-level values of the four-fermion operators are also
calculated (see Appendix \ref{Append:B}).{\footnote{We thank 
G. P. Lepage for pointing out the existence of tree-level
contributions to these terms.} They are:
\bmath {\label{Four_Fermion_coeff:1}}
\beq
    b'_{12} = \frac{g_0^2}{8},
\eeq
\beq
    b'_{14} = \frac{g_0^2}{16},
\eeq
\beq
    b'_{13} = \frac{g_0^2}{24},
\eeq
\emath
and all other $b'_i$ are zero.

\section{On-shell improvement}{\label{sect:4}}

The on-shell improved action is not unique. Given one improved
action, we can obtain another one by a
transformation of the fields. However, all these actions are
equivalent because all such actions will give the same value for a
specific on-shell quantity. Thus, we can choose to minimize the number
of operators occurring in the on-shell improved action by an
appropriate definition of the field variables.

\subsection{Isospectral transformation on fermion fields}

To simplify the improved fermion action
Eq.~(\ref{improved_fermion_action}), we consider the following
transformation:
\bmath {\label{iso_transf}}
\beq
    \chi \longrightarrow \Bigl(1 
      + a^2\varepsilon_1 m\Slash{\cal{D}}
      + a^2\varepsilon_2 \Slash{\cal{D}}^2
      + a^2\varepsilon_3 {\cal{D}}^2 \Bigr)\,\chi,
\eeq
\beq
    \bar{\chi} \longrightarrow \bar{\chi}
    	\Bigl(1 
      +\,a^2\varepsilon'_1 m\Back{\Slash{\cal{D}}}
      +\,a^2\varepsilon'_2 \Bk{\Slash{\cal{D}}}
      +\,a^2\varepsilon'_3 \Bk{\cal{D}}
        \Bigr).
\eeq
\emath
This is the most general transformation of the fermion fields $\chi$
and $\bar\chi$ which preserves their transformation properties under
the lattice symmetries.
%
After rescaling the fermion fields and redefining the fermion
mass parameter
\beq
    m' = m\bigl(1-a^2m^2(\varepsilon_1 - \varepsilon_1')),
\eeq
to the first order in the $\varepsilon_i$'s and second order in $a$,
we end up with the change of the action
\bea
    \delta S_F = a^2 \Bigl[
	(\varepsilon'_3 - \varepsilon_3){\cal{O}}_2
      +	(\varepsilon'_3 + \varepsilon_3){\cal{O}}_3
      +	(\varepsilon'_2 + \varepsilon_2
      	    + \varepsilon'_3 + \varepsilon_3){\cal{O}}_5
		\nonumber	\\
      +	(\varepsilon_1 - \varepsilon'_1
      	    + \varepsilon_2 + \varepsilon'_2){\cal{O}}_6
      +	(\varepsilon_3 + \varepsilon'_3){\cal{O}}_7
    	\Bigr] + O(a^3).
\eea
The reality of the transformed action requires that
\bmath
\beq
    \varepsilon'_3 = \varepsilon_3^\ast,
\eeq
\beq
    \varepsilon_2 + \varepsilon'_2 = {\rm real},
\eeq
{\rm and}
\beq
    \varepsilon_1 - \varepsilon'_1 = {\rm real}.
\eeq
\emath
Thus, we can always choose appropriate values of the $\varepsilon_i$'s
and $\varepsilon_i'$'s to make the coefficients $b_2, b_3, b_5, b_6$
in Eq.~(\ref{improved_fermion_action}) vanish after it is
rewritten by the new field variables.
For example,
\bmath
\beq
    \varepsilon'_3 = -\frac{1}{2}(b_3 + b_2),
\eeq
\beq
    \varepsilon_3 = -\frac{1}{2}(b_3 - b_2),
\eeq
\beq
    \varepsilon'_2 + \varepsilon_2 = b_3 - b_5,
\eeq
\beq
    \varepsilon_1 - \varepsilon'_1 = b_5 - b_3 - b_6.
\eeq
\emath
Notice that this argument is valid to any order of $g_0^2$.
Hence, to all orders in perturbation theory, we can always choose
\beq
    b_2(g_0^2, ma) =
    b_3(g_0^2, ma) =
    b_5(g_0^2, ma) =
    b_6(g_0^2, ma) = 0.
\eeq

\subsection{Isospectral transformation on gauge fields}

The general form of the gauge field transformation which changes the
action at $O(a^2)$ can be written as:
\beq{\label{gauge_transf}}
    U_\mu(x) \longrightarrow U'_\mu(x) = 
    	\exp\bigl[\epsilon X_\mu(x)\bigr] U_\mu(x).
\eeq
$U'_\mu(x)$ and $U_\mu(x)$ must have the same transformation
properties under lattice symmetry group.
The general form of $X_\mu(x)$
has been given in ref.~\cite{Luscher:2} for the case in which
$X_\mu(x)$ depends on only the gauge variables.
It is the anti-hermitian traceless part of another
field $Y_\mu(x)$:
\beq
    X_\mu(x) = Y_\mu(x) - Y_\mu(x)^\dagger -\frac{1}{N}\Tr[Y_\mu(x)
      	- Y_\mu(x)^\dagger],
\eeq
with $N = 3$ for $SU(3)$ gauge theory, and
\bea
    Y_\mu(x) = \frac{1}{4}\sum_\nu \Bigl( U_\nu(x)U_\mu(x+\hat{\nu})
    	U_\nu^\dagger(x+\hat{\mu})U_\mu^\dagger(x)
		\nonumber  \\
      - U_\mu(x)U_\nu^\dagger(x-\hat{\nu}+\hat{\mu})
        U_\mu^\dagger(x-\hat{\nu})U_\nu(x-\hat{\nu}) \Bigr).
\eea

Under the field transformation Eq.~(\ref{gauge_transf}),
the path integral is invariant
\bea
    \int [dU] \exp^{S_F[U] + S_G[U]}
    = \int [dU'] \exp^{S_F[U'] + S_G[U']}
    	\nonumber \\
    = \int [dU] \Bigl(1+\Delta J +\epsilon
        [\Delta S_G + \Delta S_F]\Bigr)
    	\exp^{S_F[U] + S_G[U]} + O(\epsilon^2).
\eea
The change of the fermion action in
Eq.~(\ref{improved_fermion_action}) is:
\beq {\label{extra_fermion_action}}
    \Delta S_F = \frac{\epsilon}{2}{\cal{O}}_4 + O(a^3).
\eeq
The jacobian $1+\Delta J$ and the change of the gauge action 
$\Delta S_G$ will not generate new terms, but only change
the coefficients $c_i$ in $S_G[U]$.
L\"uscher and Weisz~\cite{Luscher:2} discussed the changes of these
coefficients in detail.
Their results showed that we can choose an
appropriate value of $\epsilon$ to set
\beq
    c_3(g_0^2) = 0
\eeq
to all orders in perturbation theory.
From Eq.~(\ref{extra_fermion_action}), we see that if we want this
to persist, the coefficient $b_4(g_0^2, ma)$ will not be zero in
general. On the other hand, we can also set $b_4(g_0^2, ma) = 0$,
however, $c_3(g_0^2)$ will then in general be non-zero.

In order to find out the possible redundant four fermion operators,
we will generalize the argument given in ref~\cite{SW:1}.
The lattice action is rewritten in a concise form as:
\beq
    S = \sum_x [\bar\chi h_x(U) \chi + \Tr f_x(U) ].
\eeq
Consider the small change of a link variable
\bmath
\beq
    \delta U_\mu(x) = i\varepsilon_{\alpha\beta}(x\!+\!c)
    \,\Re[t^a U_\mu(x\!+\!c)]_{\alpha\beta}~t^a\,U_\mu(x),
\eeq
\beq
    \delta U^\dagger_\mu(x) = -i\varepsilon_{\alpha\beta}(x\!+\!c)
    \,\Re[t^a U_\mu(x\!+\!c)]_{\alpha\beta}~U^\dagger_\mu(x)\,t^a,
\eeq
\emath
where $\varepsilon$ is a real small number.
The jacobian differs from 1 in order $\varepsilon^2$.
To the first order of $\varepsilon$,
we get the following identity
\bmath
\bea
    \int dU_\mu(x)~e^{S[\chi,\bar\chi,U]}~\biggl\{\,\bar\chi
    t^a U_\mu(x) \frac{\partial h_x(U_\mu(x))}{\partial U_\mu(x)}
    \chi \,\varepsilon_{\alpha\beta}(x\!+\!c)\,
    \Re[ t^a U_\mu(x\!+\!c)]_{\alpha\beta}
    	\nonumber \\
    + \Tr\Bigl[t^a U_\mu(x)\frac{\partial f_x(U_\mu(x))}
    {\partial U_\mu(x)}\Bigr]
    \,\varepsilon_{\alpha\beta}(x\!+\!c)\,\Re[t^a
    U_\mu(x\!+\!c)]_{\alpha\beta}\,\biggr\} = 0.
\eea
Similarly, we get
\bea
    \int dU_\mu(x)~e^{S[\chi,\bar\chi,U]}~\biggl\{\,\bar\chi
    t^a U_\mu(x) \frac{\partial h_x(U_\mu(x))}{\partial U_\mu(x)}
    \chi \,\varepsilon_{\alpha\beta}(x\!+\!c)\,
    \Im[ t^a U_\mu(x\!+\!c)]_{\alpha\beta}
    	\nonumber \\
    + \Tr\Bigl[t^a U_\mu(x)\frac{\partial f_x(U_\mu(x))}
    {\partial U_\mu(x)}\Bigr]
    \,\varepsilon_{\alpha\beta}(x\!+\!c)\,\Im[t^a
    U_\mu(x\!+\!c)]_{\alpha\beta}\,\biggr\} = 0.
\eea
\emath
Combining the above two equations, we get
\bea
    Q^a_{\alpha\beta}(\bar\chi, \chi) =
    \int dU_\mu(x)~e^{S[\chi,\bar\chi,U]}~\biggl\{\,\bar\chi
    t^a U_\mu(x) \frac{\partial h_x(U_\mu(x))}{\partial U_\mu(x)}
    \chi \,\varepsilon_{\alpha\beta}(x\!+\!c)\,
    [t^a U_\mu(x\!+\!c)]_{\alpha\beta}
    	\nonumber \\
    + \Tr\Bigl[t^a U_\mu(x)\frac{\partial f_x(U_\mu(x))}
    {\partial U_\mu(x)}\Bigr]
    \,\varepsilon_{\alpha\beta}(x\!+\!c)\,[t^a
    U_\mu(x\!+\!c)]_{\alpha\beta}\,\biggr\} = 0.
\eea

In the above equation,
we replace $\varepsilon_{\alpha\beta}(x\!+\!c)$
by $a^3\bar\chi^\alpha(x\!+\!c)\chi^\beta(x\!+\!c\!+\!\mu)$,
multiply it by a combined phase factor
\beq
    \sum_{i=11}^{18}\,\epsilon'_i\eta_\mu(x) P_\mu^{i}(c),
\eeq
and sum it on the hypercubic vector $c$.
After substituting $h_x(U)$ and $f_x(U)$ by
the actual staggered fermion action and the gauge action,
we see that we can add the following terms
\beq {\label{Delta_S:2}}
    \Delta S = \frac{a^2}{2}\sum_{i=11}^{18}\,\epsilon'_i
        {\cal F}_i + a^2 \epsilon'_{11} {\cal O}_4 + O(a^3)
\eeq
to the action without changing the path integral to the order
of $O(a^3)$. 
Notice that because of the identity
\beq
    \sum_c \eta_\mu(c) P_\mu^{i}(c) = \delta_{i,11},
    \qquad i = 11,\cdots,18
\eeq
there is only one dimension-6 bilinear in Eq.~(\ref{Delta_S:2}).
Therefore, we conclude that the four fermion operators
${\cal F}_{11} \cdots {\cal F}_{18}$,
whose fermion bilinears consist of two sites separated by
one link, are redundant and
their coupling constants $b'_{11} \cdots b'_{18}$ can be set
to be zero. Again, the coefficient of operator
${\cal O}_4$ will get a change accordingly.
Is it possible to get rid of some other four fermion operators?
From the above discussion, we conclude ``no''. Because in the
original staggered fermion action, $\chi$ and $\bar\chi$ are
separated by one link, there is no way to generate an operator
like ${\cal F}_1, \cdots, {\cal F}_{10}$ when we multiply that link
variable by some Grassmann variables.

\subsection{$O(a^2)$ on-shell improved action}

The $O(a^2)$ improved action for lattice QCD can be
written as:
\beq
    S_{QCD} = S_G[U] + S_F[\chi, \bar\chi, U],
\eeq
where $S_F$ is given by Eq.~(\ref{improved_fermion_action}) and
$S_G$ is given by Eq.~(\ref{improved_gauge_action}).

The operator ${\cal{O}}_4$ involves the lattice sites in two
nearest neighbor hypercubes. There exists another operator
${\cal{O}}'_4$ which only involves lattice sites inside
one hypercube and is equivalent to ${\cal{O}}_4$ up to an
operator of dimension of seven. We can construct this operator by
replacing the link variable $U_\mu(x)$ in the Dirac
operator ~$\Slash{\cal{D}}$ with a modified link (e.\,g.~the MILC
``fat link'') ${\cal{W}}_\mu(x)U_\mu(x)$~\cite{MILC:2}:
\beq
    {\cal{O}}'_4 = \sum_{x,\mu} \bar{\chi}(x) \frac{\eta_\mu(x)}{2a}
    	\left[ {\cal{W}}_\mu(x)U_\mu(x) \chi(x+\hat{\mu})
	- U_\mu^\dagger(x-\hat{\mu}){\cal{W}}_\mu^\dagger(x-\hat{\mu})
	        \chi(x-\hat{\mu}) \right].
\eeq
This new operator ${\cal{O}}'_4$ obeys all the staggered fermion
lattice symmetries.
The factor ${\cal{W}}_\mu(x)$ has the form:
\bea
    {\cal{W}}_\mu(x) = \frac{1}{a^2}\,\sum_{\nu \neq \mu} \Bigl[
    	U_\nu(x)U_\mu(x+\hat{\nu})U_\nu^\dagger(x+\hat{\mu})
	U_\mu^\dagger(x)  \nonumber  \\
      + U_\nu^\dagger(x-\hat{\nu})U_\mu(x-\hat{\nu})
	U_\nu(x+\hat{\mu}-\hat{\nu})U_\mu^\dagger(x) - 2 \Bigr],
\eea
and its continuum limit is $ig_0a[D_\nu, F_{\nu\mu}]$.
%
Thus, ${\cal{O}}'_4$ has the same continuum
limit as ${\cal{O}}_4$.
Because it is simpler, we will choose ${\cal{O}}'_4$
instead of ${\cal{O}}_4$ to make our on-shell improved
action.

From the above discussion, we can see that one possible on-shell
improved action for lattice QCD can be constructed as
\beq {\label{action:1}}
    S_{QCD}^{(1)} = \sum_{i=0}^2 c_i(g_0^2) {\cal{L}}_i
    	+ S_F^{(1)},
\eeq
with
\bea {\label{fermion_action:1}}
    S_F^{(1)} = \bar{\chi}(\Slash{\cal{D}} + m)\chi
	+ a^2 b_1(g_0^2, ma) {\cal{O}}_1
	+ a^2 b_4(g_0^2, ma) {\cal{O}}'_4
	+ a^2 b_7(g_0^2, ma) {\cal{O}}_7 \nonumber \\
	+ a^2 \sum_{i=1}^{10}b'_i((g_0^2, ma) {\cal{F}}_i.
\eea
Another possible choice would be:
\beq {\label{action:2}}
    S_{QCD}^{(2)} = \sum_{i=0}^3 c_i(g_0^2) {\cal{L}}_i
    	+ S_F^{(2)},
\eeq
with
\beq {\label{fermion_action:2}}
    S_F^{(2)} = \bar{\chi}(\Slash{\cal{D}} + m)\chi
	+ a^2 b_1(g_0^2, ma) {\cal{O}}_1
	+ a^2 b_7(g_0^2, ma) {\cal{O}}_7
	+ a^2 \sum_{i=1}^{10}b'_i((g_0^2, ma) {\cal{F}}_i.
\eeq
In either case, the tree-level improved action is the same because
both $b_4(0, ma)$ and $c_3(0)$ vanish.

\subsection{Formulae arranged for lattice computation}

For the operators appearing in Eq.~(\ref{action:1}) or
Eq.~(\ref{action:2}), the gauge part,
given by Eq.~(\ref{improved_gauge_action}), is already in a
form handy for numerical simulation. However, the fermion bilinears
in $S_F^{(1)}$ or $S_F^{(2)}$ are represented in terms of the
hypercubic fields. In order to do numerical calculation, it is
convenient to rewrite them in terms of the
original field variables $\chi$ and $\bar{\chi}$.

Let us first consider the four fermion operators. These 10
operators, described in section \ref{sect:2} and \ref{sect:4}, are
already written in terms of the original variables $\chi$ and
$\bar\chi$. However, they can be simulated using known Monte Carlo
techniques only if auxiliary Yukawa fields are introduced so these
four fermion operators can be written in a bilinear form. The
resulting gauge-Yukawa fermion action will be quite complicated with
hermiticity properties that depend on the sign of the original four
fermion coefficients. Further, the positive-definite character of
the staggered fermion action may be lost unless these new Yukawa
terms possess some additional (staggered fermion) $U_A(1)$ 
symmetry. Since these four fermion terms are not present in tree
level (or tadpole improved tree-level) approximation, we will not
consider this question further in this paper.

Only considering the fermion bilinears,
the fermion action $S_F^{(1)}$ in Eq.~(\ref{fermion_action:1})
can be rewritten as:
\bea {\label{new_fermion:1}}
    S_F^{(1)} = a^4\sum_{x,\mu} \bar{\chi}(x)
	\frac{\eta_\mu(x)}{2a} \Bigl[
	    {\cal{U}}_\mu(x)\chi(x+\hat\mu) -
	    {\cal{U}}_\mu^\dagger(x-\hat\mu)\chi(x-\hat\mu) \Bigr]
		\nonumber \\
	+ \bigl(1-\alpha_3(g_0^2,ma)\bigr)m~
	  a^4 \sum_x \bar{\chi}(x)\chi(x)
		\nonumber \\
	+ \alpha_3(g_0^2, ma)m~ a^4\sum_{x,\mu}
	    \bar{\chi}(x)\frac{1}{2}\Bigl[
	    U(x, x+2\hat\mu)\chi(x+2\hat\mu) +
	    U(x, x-2\hat\mu)\chi(x-2\hat\mu) \Bigr]
		\nonumber \\
	- \alpha_1(g_0^2, ma)~ a^4\sum_{x,\mu}
	    \bar{\chi}(x)\frac{\eta_\mu(x)}{6a} \Bigl[
	    U(x, x+3\hat\mu)\chi(x+3\hat\mu) -
	    U(x, x-3\hat\mu)\chi(x-3\hat\mu) \Bigr],
\eea
with
\bmath
\beq {\label{new_U:1}}
    {\cal{U}}_\mu(x) = \bigl(1+\alpha_1(g_0^2,ma)-\alpha_2(g_0^2,ma)
	\bigr)U_\mu(x) + \alpha_2(g_0^2,ma)\tilde{U}_\mu(x),
\eeq
\beq
    \tilde{U}_\mu(x) = \frac{1}{6} \sum_{\stackrel{\nu}{\nu\neq\mu}}
    	\Bigl[U_\nu(x)U_\mu(x+\hat\nu)U_\nu^\dagger(x+\hat\mu)
	+ U_\nu^\dagger(x-\hat\nu)U_\mu(x-\hat\nu)U_\nu(x-\hat\nu
	  +\hat\mu) \Bigr],
\eeq
\emath
and
\bmath
\beq
    U(x,x+2\hat\mu) = U_\mu(x)U_\mu(x+\hat\mu),
\eeq
\beq
    U(x,x-2\hat\mu) = U_\mu^\dagger(x-\hat\mu)U_\mu^\dagger(x
    	-2\hat\mu),
\eeq
\beq
    U(x,x+3\hat\mu) = U_\mu(x)U_\mu(x+\hat\mu)U_\mu(x+2\hat\mu),
\eeq
\beq
    U(x,x-3\hat\mu) = U_\mu^\dagger(x-\hat\mu)U_\mu^\dagger(x
    	-2\hat\mu)U_\mu^\dagger(x-3\hat\mu).
\eeq
\emath
The three parameters $\alpha_1, \alpha_2$ and $\alpha_3$ are real
numbers and related to the parameters $b_1, b_4$ and $b_7$ of
Eq.~(\ref{fermion_action:1}) by:
\bmath
\beq
    \alpha_1 = -\frac{3}{4}b_1,
\eeq
\beq
    \alpha_2 = b_4,
\eeq
\beq
    \alpha_3 = \frac{1}{2}b_7,
\eeq
\emath

The form of $S_F^{(2)}$, when rewritten using the fermion fields
$\chi$ and $\bar{\chi}$, is the same as Eq.~(\ref{new_fermion:1})
but with $\alpha_2(g_0^2,ma) = 0$:
\beq
    F_F^{(2)} = S_F^{(1)}\bigl|_{\alpha_2(g_0^2, ma) = 0}.
\eeq

At the tree-level, we have
\beq
   \alpha_1(0, ma) = \frac{1}{8}, \qquad
   \alpha_2(0, ma) = \alpha_3(0, ma) = 0.
\eeq
One way to do a better job than tree-level improvement
may be to use tadpole improvement.
Following the work of Lepage and Mackenzie~\cite{Lepage:1}, we can
replace all links $U_\mu(x)$ which appear in lattice operators by
$\frac{1}{u_0}U_\mu(x)$ ~(Of course, the normalization of the
coupling constant will be different).
The constant $u_0$ is the mean value of
the link matrix and is measured in the simulation by the quantity
$u_0 = \left[\Re\langle \frac{1}{3}\Tr U_\Box \rangle
\right]^\frac{1}{4}$.
The tadpole improved tree level coefficients are the same as above
except that the parameter $\alpha_1$ which appears in 
Eq.~(\ref{new_fermion:1}) and Eq.~(\ref{new_U:1}) must be replaced
by two different parameters, $\alpha_1^a$ and $\alpha_1^b$.
The value of the parameter $\alpha_1^a$ appearing in
Eq.~(\ref{new_fermion:1}) is $\alpha_1 = 1/8u_0^2$ while
the parameter $\alpha_1^b$ in Eq.~(\ref{new_U:1}) is 
still 1/8. \footnote{Thanks
Attila Mihaly for pointing out this.}

\section{Conclusion}{\label{sect:5}}

Using Symanzik's improvement program, we have discussed the $O(a^2)$
on-shell improvement of the staggered fermion action in a
systematic way. 
Our first step was to find all dimension-6 lattice operators
which are scalars under the lattice symmetry group.
We found that there are 5 linearly independent
fermion bilinears that are invariant under all lattice symmetry
transformations. When the mass parameter is not zero, the $U_A(1)$
symmetry of the staggered fermion is violated, and there are 2 more
fermion bilinears that violate only this $U_A(1)$ symmetry and are
proportional to the mass of the fermions. For staggered fermions, we
observed that there are 18 independent four fermion operators.
Therefore, we have at most 25 fermion operators which can be added as
counterterms to the standard staggered fermion
action to remove all $O(a^2)$ errors from all physical quantities. 
Including the 3 independent dimension-6 gauge operators, we end up
with 28 counterterms for the $O(a^2)$ improved lattice QCD with
staggered fermions.

For on-shell improvement, we can use the isospectral transformation
of the field variables to eliminate all possible redundant
operators. Including such field transformations,
we concluded that we need at most 15 independent lattice operators
of dimension-6 to construct the $O(a^2)$ on-shell improved
lattice QCD with staggered fermions.
Ten of these are four-fermion operators,
which are absent at the tree-level, and hence of the order of
$O(g_0^4a^2)$ at most.
The other 5 are fermion bilinears and gauge operators and
only two of them are nonzero at tree-level.
Two possible improved actions are given by 
Eq.~(\ref{action:1}) and Eq.~(\ref{action:2}).

Thus, we found that the Naik term
is not the only term in the improved staggered fermion action.
It is worth emphasizing that to remove the $O(a^2)$ errors
from lattice computation, we must use both an improved gauge
action and an improved fermion action at the same time, not just
one of them.

The recent numerical results from the 
MILC~\cite{MILC:1,MILC:2}
and Bielefeld~\cite{Bielefeld:1}
groups are consistent with our analysis.
Furthermore, in the free case, our result is the same as Naik's,
as should be expected given the equivalence of free
Dirac-K\"ahler and staggered fermions.

\bigskip
\bigskip

\acknowledgements

I would like to thank Professor Norman H. Christ. Without our
intensive discussions during all stages of this work, this paper
could not have been finished.
This research was supported by the U. S. Department of Energy under
grant DE-FG02-92 ER40699.

\appendix
\begin{appendix}
\section{Classical continuum limit of staggered fermion action}
{\label{Append:A}}

In this appendix, we will evaluate the continuum limit of
the staggered fermion action,
\beq \label{standard_stag}
    S_F = a^4\sum_x \bar{\chi}(x) \Bigl[ \sum_\mu \eta_\mu(x)
    	\frac{1}{2a}\Bigl( U_\mu(x) \chi(x + \hat{\mu}) -
	U_\mu^{\dagger}(x - \hat{\mu}) \chi(x - \hat{\mu}) \Bigr)
	+ m \chi(x) \Bigr],
\eeq
through order $a^2$.
As the first step in this derivation, we rewrite this action
using the covariant hypercubic fermion fields defined in 
Eq.~(\ref{cov_field}) in which ${\cal{U}}_A(y)$, 
the average of the link products along the shortest paths from
site $y$ to site $y+A$, is defined as:
\beq
    {\cal{U}}_A(y) = \frac{1}{4!}\sum_{P_{(\mu\nu\rho\sigma)}}
    	U_\mu(y)^{A_\mu} U_\nu(y\!+\!A_\mu \hat{\mu})^{A_\nu}
	U_\rho(y\!+\!A_\mu\hat{\mu}\!+\!A_\nu\hat{\nu})^{A_\rho}
	U_\sigma(y\!+\!A_\mu\hat{\mu}\!+\!A_\nu
	    \hat{\nu}\!+\!A_\rho\hat{\rho})^{A_\sigma},
\eeq
where the summation is on all permutations of $(\mu\nu\rho\sigma)$.
We define the hypercubic first and second order covariant
derivatives as
\bmath {\label{cov_derivatives}}
\beq {\label{cov_derivative:D}}
    \overline{D}_\mu \varphi_A(y) = \frac{1}{4a} \Bigl[
    	U_\mu(y)U_\mu(y+\hat{\mu})\varphi_A(y+2\hat{\mu})
	-U_\mu^\dagger(y-\hat{\mu})U_\mu^\dagger(y-2\hat{\mu})
	\varphi_A(y-2\hat{\mu}) \Bigr],
\eeq
\beq
    \overline{\Delta}_\mu\varphi_A(y) = \frac{1}{4a^2} \Big[
    	U_\mu(y)U_\mu(y+\hat{\mu})\varphi_A(y+2\hat{\mu})
	+U_\mu^\dagger(y-\hat{\mu})U_\mu^\dagger(y-2\hat{\mu})
	\varphi_A(y-2\hat{\mu}) -2\varphi_A(y) \Bigr].
\eeq
\emath
Then, with no approximation, we can rewrite
Eq.~(\ref{standard_stag}) as
\bea
    S_F = (2a)^4 \sum_y \sum_{AB}
       \sum_\mu ~\bar{\varphi}_A(y)\biggl\{
    	\overline{(\gamma_\mu \otimes I)}_{AB}~\frac{1}{2}
	    \Bigl[w_\mu^{(1)}(y;AB) + w_\mu^{(2)}(y;AB)
	    \Bigr] \overline{D}_\mu
	    \nonumber       \\
       -\overline{(\gamma_5\otimes \xi_{5\mu})}_{AB}~\frac{1}{2}
	    \Bigl[w_\mu^{(1)}(y;AB) - w_\mu^{(2)}(y;AB)
	    \Bigr] \overline{D}_\mu
	    \nonumber	\\
       +\overline{(\gamma_\mu \otimes I)}_{AB}~\frac{a}{2}
            \Bigl[w_\mu^{(1)}(y;AB) - w_\mu^{(2)}(y;AB)
	    \Bigr] \overline{\Delta}_\mu
	    \nonumber       \\
       -\overline{(\gamma_5\otimes \xi_{5\mu})}_{AB}~\frac{a}{2}
            \Bigl[w_\mu^{(1)}(y;AB) + w_\mu^{(2)}(y;AB)
	    \Bigr] \overline{\Delta}_\mu
	    \nonumber       \\
       +\overline{(\gamma_\mu \otimes I)}_{AB}~\frac{1}{4a}
            \Bigl[w_\mu^{(1)}(y;AB) + w_\mu^{(4)}(y;AB)
	    -w_\mu^{(2)}(y;AB) -w_\mu^{(3)}(y;AB) \Bigr]
	    \nonumber       \\
       -\overline{(\gamma_5\otimes \xi_{5\mu})}_{AB}~\frac{1}{4a}
            \Bigl[w_\mu^{(1)}(y;AB) + w_\mu^{(2)}(y;AB)
	    -w_\mu^{(3)}(y;AB) -w_\mu^{(4)}(y;AB) \Bigr]
	    \biggr\} \varphi_B(y)
	    \nonumber       \\
      + (2a)^4 \sum_y \sum_A~m\bar{\varphi}_A(y)\varphi_A(y).
\eea
The two hypercubic vectors $A$ and $B$ satisfy the
delta function $\bar\delta(A\!+\!B\!+\!\hat\mu)$, and the ``bar''
means modulo 2.
The four closed-loop link products are defined as:
\bmath
\beq
    w_\mu^{(1)}(y;AB) = \delta_{A_\mu 1}\,{\cal{U}}_A(y)
    	U_\mu(y\!+\!A)\,{\cal{U}}_B^\dagger(y\!+\!2\hat\mu)
	U_\mu^\dagger(y\!+\!\hat\mu)U_\mu^\dagger(y),
\eeq
\beq
    w_\mu^{(2)}(y;AB) = \delta_{A_\mu 0}\,{\cal{U}}_A(y)
    	U_\mu^\dagger(y\!-\!2\hat\mu\!+\!B)\,
	{\cal{U}}_B^\dagger(y\!-\!2\hat\mu)
	U_\mu(y\!-\!2\hat\mu)U_\mu(y\!-\!\mu),
\eeq
\beq
    w_\mu^{(3)}(y;AB) = \delta_{A_\mu 1}\,{\cal{U}}_A(y)
    	U_\mu^\dagger(y\!+\!B)\,{\cal{U}}_B^\dagger(y),
\eeq
\beq
    w_\mu^{(4)}(y;AB) = \delta_{A_\mu 0}\,{\cal{U}}_A(y)
    	U_\mu(y\!+\!A)\,{\cal{U}}_B^\dagger(y).
\eeq
\emath
If we represent ${\cal{U}}_A(y)$ by a doubled line
which starts at site $y$ and ends at site $y+A$, and 
${\cal{U}}_A^\dagger(y)$ by a doubled line
which starts at site $y+A$ and ends at site $y$,
then some typical loops can be shown
by four figures(Figure \ref{fig:loop1} to \ref{fig:loop4}).

To expand the links in powers of ``$a$'', we take advantage of the
parallel transporter from $x$ to $x + \hat{\mu}$ to define the
gauge field ${\cal{A}}_\mu(x)$ by the path ordered exponential
\beq
    U_\mu(x) = P\exp\biggl\{ig_0a\int_0^1 d\tau {\cal{A}}_\mu(
    	x + \tau\hat{\mu}) \biggr\}.
\eeq

When $a \rightarrow 0$, we have
\bmath
\beq
    \overline{D}_\mu = D_\mu + \frac{2}{3}a^2 D_\mu^3 + O(a^3),
\eeq
\beq
    \overline{\Delta}_\mu = D_\mu^2 + O(a^2).
\eeq
\emath
where $D_\mu = \partial_\mu + ig_0{\cal{A}}_\mu$ is the continuum
covariant derivative.
Expanding the $w$'s in powers of ``$a$'', we get
\bmath
\beq
    \frac{1}{2}\bigl[w_\mu^{(1)} + w_\mu^{(2)}\bigr] = 1 + O(a^3),
\eeq
\beq
    \frac{1}{2}\bigl[w_\mu^{(1)} - w_\mu^{(2)}\bigr] = 
    	\frac{3}{2}ig_0a^2\sum_\nu A_\nu F_{\nu\mu} + O(a^3),
\eeq
\beq
    \frac{1}{4a}\bigl[w_\mu^{(1)} + w_\mu^{(4)} - w_\mu^{(2)}
        -w_\mu^{(3)}\bigr] = ig_0a\sum_\nu A_\nu F_{\nu\mu}
	+\frac{1}{2}ig_0a^2\sum_{\lambda\nu}A_\lambda A_\nu
	\left[D_\nu,F_{\lambda\mu}\right] + O(a^3),
\eeq
\beq
    \frac{1}{4a}\bigl[w_\mu^{(1)} + w_\mu^{(2)} - w_\mu^{(3)}
        -w_\mu^{(4)}\bigr] = \frac{1}{2}ig_0a^2\sum_\nu A_\nu
	\left[D_\mu,F_{\nu\mu}\right] + O(a^3).
\eeq
\emath
Finally, we get the classical continuum limit of the staggered
fermion action as Eq.~(\ref{classical_limit}).

\section{Calculation of the tree-level coefficients of four-fermion
operators}{\label{Append:B}}
We evaluate the amplitude represented by the graph shown in
Figure \ref{fig:feynman} for the case of vanishing external momenta,
$p_i^{\mu} = 0$.
The amplitude is given by:
\bea
    K_{ABCD}^{abcd} = -g_0^2 t^i_{ab} t^i_{cd} \sum_\mu \sum_M{\!}^{'}
    \frac{1}{\hat\pi_M^2}\cos{[\frac{1}{2}(\pi_A\!+\!\pi_B)_\mu]}
    \delta(A\!+\!B\!+\!M\!+\!\hat\eta_\mu)
    	\nonumber \\
    \cos{[\frac{1}{2}(\pi_C\!+\!\pi_D)_\mu]}
    \delta(C\!+\!D\!+\!M\!+\!\hat\eta_\mu).
\eea
where $a,b,c$, and $d$ are color indices, and
$\hat\eta_\mu$ is a hypercubic vector whose $\nu$'s component is 1 
only if $\nu < \mu$.
$\pi_M$ is the momentum propagated
by the gluon and  the quantity $\hat\pi_M^2$ is defined as
\beq
    \hat\pi_M^2 = \frac{4}{a^2}\sum_\nu\sin^2(M_\nu\pi/2).
\eeq
The primed summation on $M$ is for all hypercubic
vectors $M$ with $M_\mu = 0$.

After some algebra and using the notation of
ref~\cite{staggered:1}, we get
\bea
    K_{ABCD}^{abcd} = -g_0^2 t^i_{ab} t^i_{cd} \Bigl\{
        \frac{a^2}{4} \sum_{\mu\neq\nu\neq\lambda}
	\overline{\overline{(\gamma_{\mu\nu}\otimes
	\xi_{5\lambda})}}_{AB}\,
	\overline{\overline{(\gamma_{\mu\nu}\otimes
	\xi_{5\lambda})}}_{CD} +
		\nonumber \\
        \frac{a^2}{8} \sum_{\mu\neq\nu\neq\lambda}
	\overline{\overline{(\gamma_{5\lambda}\otimes
	\xi_{\mu\nu})}}_{AB}\,
	\overline{\overline{(\gamma_{5\lambda}\otimes
	\xi_{\mu\nu})}}_{CD} +
		\nonumber \\
        \frac{a^2}{12} \sum_\mu
	\overline{\overline{(I\otimes
	\xi_\mu)}}_{AB}\,
	\overline{\overline{(I\otimes
	\xi_\mu)}}_{CD}
    \Bigr\},
\eea
with
\beq
    \overline{\overline{(\gamma_S\otimes\xi_F)}}_{AB} =
    \sum_{CD} \frac{1}{4}(-1)^{A\cdot C}\,
    \overline{(\gamma_S\otimes\xi_F)}_{CD}\,
    \frac{1}{4} (-1)^{D\cdot B}.
\eeq
Then the action differs from the continuum by a four fermion term:
\beq
    \Delta S = \frac{1}{2}\sum_{ABCD}\sum_{abcd}
    K_{ABCD}^{abcd}\bar\chi^a(\pi_A)\chi^b(\pi_B)
    \bar\chi^c(\pi_C)\chi^d(\pi_D).
\eeq
Written in terms of the hypercubic fields, we get
\beq
    \Delta S = -\frac{g_0^2}{2} \bigl(
        \frac{a^2}{4}{\cal F}_{12} + \frac{a^2}{8}{\cal F}_{14}
	+\frac{a^2}{12}{\cal F}_{13}
    \bigr).
\eeq
The counterterm is the opposite of the above term. Hence we get
the coefficients listed in Eq.~(\ref{Four_Fermion_coeff:1}).

\end{appendix}

\pagebreak[4]

\begin{figure}[htbp]
  \setlength{\unitlength}{0.012500in}%
\begingroup\makeatletter\ifx\SetFigFont\undefined
\def\x#1#2#3#4#5#6#7\relax{\def\x{#1#2#3#4#5#6}}%
\expandafter\x\fmtname xxxxxx\relax \def\y{splain}%
\ifx\x\y   
\gdef\SetFigFont#1#2#3{%
  \ifnum #1<17\tiny\else \ifnum #1<20\small\else
  \ifnum #1<24\normalsize\else \ifnum #1<29\large\else
  \ifnum #1<34\Large\else \ifnum #1<41\LARGE\else
     \huge\fi\fi\fi\fi\fi\fi
  \csname #3\endcsname}%
\else
\gdef\SetFigFont#1#2#3{\begingroup
  \count@#1\relax \ifnum 25<\count@\count@25\fi
  \def\x{\endgroup\@setsize\SetFigFont{#2pt}}%
  \expandafter\x
    \csname \romannumeral\the\count@ pt\expandafter\endcsname
    \csname @\romannumeral\the\count@ pt\endcsname
  \csname #3\endcsname}%
\fi
\fi\endgroup
\begin{picture}(166,125)(57,660)
\thinlines
\put(140,759){\circle*{6}}
\put(220,760){\circle*{6}}
\put(220,680){\circle*{6}}
\put(140,680){\circle*{6}}
\put( 60,680){\circle*{6}}
\thicklines
\put(140,760){\line( 1, 0){ 80}}
\put( 60,680){\line( 1, 0){160}}
\put(170,760){\vector( 1, 0){ 15}}
\put(185,680){\vector(-1, 0){ 15}}
\put(110,680){\vector(-1, 0){ 20}}
\special{ps: gsave 0 0 0 setrgbcolor}\put( 58,680){\line( 1, 1){ 80}}
\special{ps: grestore}\special{ps: gsave 0 0 0 setrgbcolor}\put(219,760){\line( 0,-1){ 80}}
\special{ps: grestore}\special{ps: gsave 0 0 0 setrgbcolor}\put(222,760){\line( 0,-1){ 80}}
\special{ps: grestore}\special{ps: gsave 0 0 0 setrgbcolor}\put(219,730){\vector( 0,-1){ 19}}
\special{ps: grestore}\special{ps: gsave 0 0 0 setrgbcolor}\put(222,730){\vector( 0,-1){ 19}}
\special{ps: grestore}\special{ps: gsave 0 0 0 setrgbcolor}\put( 92,710){\vector( 1, 1){ 17}}
\special{ps: grestore}\special{ps: gsave 0 0 0 setrgbcolor}\put( 61,679){\line( 1, 1){ 80}}
\special{ps: grestore}\special{ps: gsave 0 0 0 setrgbcolor}\put( 90,712){\vector( 1, 1){ 17}}
\special{ps: grestore}\put( 60,660){\makebox(0,0)[lb]{\smash{\SetFigFont{12}{14.4}{rm}y}}}
\put(210,660){\makebox(0,0)[lb]{\smash{\SetFigFont{12}{14.4}
{rm}y+2$\hat\mu$}}}
\put(210,770){\makebox(0,0)[lb]{\smash{\SetFigFont{12}{14.4}
{rm}y+2$\hat\mu$+B}}}
\put(135,770){\makebox(0,0)[lb]{\smash{\SetFigFont{12}{14.4}{rm}y+A}}}
\end{picture}
  \vspace{0.6cm}
  \caption{\label{fig:loop1}A typical graph of $w_\mu^{(1)}(y;AB)$.
  	It starts from $y$, through $y\!+\!A$,
	$y\!+\!2\mu\!+\!B$, $y\!+\!2\mu$,
	$y\!+\!\mu$, and then comes back to $y$.
	The two hypercubic vectors $A$ and $B$ satisfy the condition
	$A_\mu=1$, $B_\mu=0$.}
\end{figure}
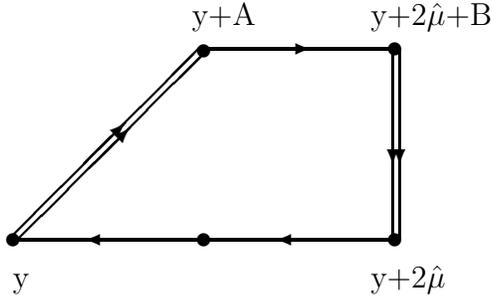

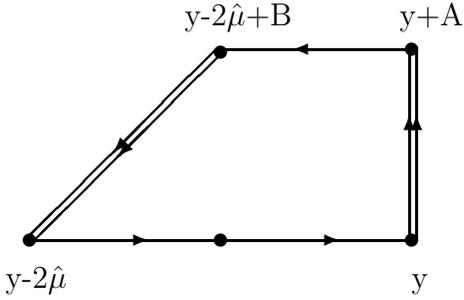
\begin{figure}[htbp]
  \setlength{\unitlength}{0.012500in}%
\begingroup\makeatletter\ifx\SetFigFont\undefined
\def\x#1#2#3#4#5#6#7\relax{\def\x{#1#2#3#4#5#6}}%
\expandafter\x\fmtname xxxxxx\relax \def\y{splain}%
\ifx\x\y   
\gdef\SetFigFont#1#2#3{%
  \ifnum #1<17\tiny\else \ifnum #1<20\small\else
  \ifnum #1<24\normalsize\else \ifnum #1<29\large\else
  \ifnum #1<34\Large\else \ifnum #1<41\LARGE\else
     \huge\fi\fi\fi\fi\fi\fi
  \csname #3\endcsname}%
\else
\gdef\SetFigFont#1#2#3{\begingroup
  \count@#1\relax \ifnum 25<\count@\count@25\fi
  \def\x{\endgroup\@setsize\SetFigFont{#2pt}}%
  \expandafter\x
    \csname \romannumeral\the\count@ pt\expandafter\endcsname
    \csname @\romannumeral\the\count@ pt\endcsname
  \csname #3\endcsname}%
\fi
\fi\endgroup
\begin{picture}(173,125)(50,660)
\thinlines
\put(140,759){\circle*{6}}
\put(220,760){\circle*{6}}
\put(220,680){\circle*{6}}
\put(140,680){\circle*{6}}
\put( 60,680){\circle*{6}}
\thicklines
\put( 60,680){\line( 1, 0){160}}
\put(140,760){\line( 1, 0){ 80}}
\put(190,760){\vector(-1, 0){ 20}}
\put( 90,680){\vector( 1, 0){ 20}}
\put(170,680){\vector( 1, 0){ 20}}
\special{ps: gsave 0 0 0 setrgbcolor}\put(138,760){\line(-1,-1){ 80}}
\special{ps: grestore}\special{ps: gsave 0 0 0 setrgbcolor}\put(219,680){\line( 0, 1){ 81}}
\special{ps: grestore}\special{ps: gsave 0 0 0 setrgbcolor}\put(222,680){\line( 0, 1){ 81}}
\special{ps: grestore}\special{ps: gsave 0 0 0 setrgbcolor}\put(219,709){\vector( 0, 1){ 24}}
\special{ps: grestore}\special{ps: gsave 0 0 0 setrgbcolor}\put(222,709){\vector( 0, 1){ 24}}
\special{ps: grestore}\special{ps: gsave 0 0 0 setrgbcolor}\put(142,760){\line(-1,-1){ 80}}
\special{ps: grestore}\special{ps: gsave 0 0 0 setrgbcolor}\put(108,730){\vector(-1,-1){ 13}}
\special{ps: grestore}\special{ps: gsave 0 0 0 setrgbcolor}\put(110,728){\vector(-1,-1){ 13}}
\special{ps: grestore}\put(220,660){\makebox(0,0)[lb]{\smash{\SetFigFont{12}{14.4}{rm}y}}}
\put(215,770){\makebox(0,0)[lb]{\smash{\SetFigFont{12}{14.4}{rm}y+A}}}
\put( 50,660){\makebox(0,0)[lb]{\smash{\SetFigFont{12}{14.4}
{rm}y-2$\hat\mu$}}}
\put(125,770){\makebox(0,0)[lb]{\smash{\SetFigFont{12}{14.4}
{rm}y-2$\hat\mu$+B}}}
\end{picture}
  \vspace{0.6cm}
  \caption{\label{fig:loop2}A typical graph of $w_\mu^{(2)}(y;AB)$.
  	It starts from $y$, through $y\!+\!A$,
  	$y\!-\!2\mu\!+\!B$, $y\!-\!2\mu$, and then comes back to $y$.
	The two hypercubic vectors $A$ and $B$ satisfy the condition
	$A_\mu=0$, $B_\mu=1$.}
\end{figure}

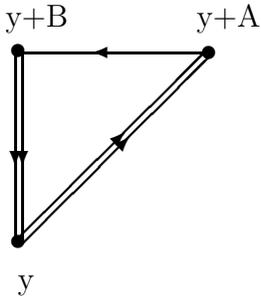
\begin{figure}[htbp]
  \setlength{\unitlength}{0.012500in}%
\begingroup\makeatletter\ifx\SetFigFont\undefined
\def\x#1#2#3#4#5#6#7\relax{\def\x{#1#2#3#4#5#6}}%
\expandafter\x\fmtname xxxxxx\relax \def\y{splain}%
\ifx\x\y   
\gdef\SetFigFont#1#2#3{%
  \ifnum #1<17\tiny\else \ifnum #1<20\small\else
  \ifnum #1<24\normalsize\else \ifnum #1<29\large\else
  \ifnum #1<34\Large\else \ifnum #1<41\LARGE\else
     \huge\fi\fi\fi\fi\fi\fi
  \csname #3\endcsname}%
\else
\gdef\SetFigFont#1#2#3{\begingroup
  \count@#1\relax \ifnum 25<\count@\count@25\fi
  \def\x{\endgroup\@setsize\SetFigFont{#2pt}}%
  \expandafter\x
    \csname \romannumeral\the\count@ pt\expandafter\endcsname
    \csname @\romannumeral\the\count@ pt\endcsname
  \csname #3\endcsname}%
\fi
\fi\endgroup
\begin{picture}(88,125)(55,660)
\thinlines
\put(140,759){\circle*{6}}
\put( 60,680){\circle*{6}}
\put( 60,760){\circle*{6}}
\thicklines
\put( 60,759){\line( 1, 0){ 80}}
\put(107,759){\vector(-1, 0){ 16}}
\special{ps: gsave 0 0 0 setrgbcolor}\put( 60,681){\line( 1, 1){ 78}}
\special{ps: grestore}\special{ps: gsave 0 0 0 setrgbcolor}\put( 61,678){\line( 1, 1){ 81}}
\special{ps: grestore}\special{ps: gsave 0 0 0 setrgbcolor}\put( 59,760){\line( 0,-1){ 81}}
\special{ps: grestore}\special{ps: gsave 0 0 0 setrgbcolor}\put( 62,760){\line( 0,-1){ 80}}
\special{ps: grestore}\special{ps: gsave 0 0 0 setrgbcolor}\put( 94,715){\vector( 1, 1){ 11}}
\special{ps: grestore}\special{ps: gsave 0 0 0 setrgbcolor}\put( 96,713){\vector( 1, 1){ 11}}
\special{ps: grestore}\special{ps: gsave 0 0 0 setrgbcolor}\put( 59,729){\vector( 0,-1){ 18}}
\special{ps: grestore}\special{ps: gsave 0 0 0 setrgbcolor}\put( 62,729){\vector( 0,-1){ 18}}
\special{ps: grestore}\put( 60,660){\makebox(0,0)[lb]{\smash{\SetFigFont{12}{14.4}{rm}y}}}
\put(135,770){\makebox(0,0)[lb]{\smash{\SetFigFont{12}{14.4}{rm}y+A}}}
\put( 55,770){\makebox(0,0)[lb]{\smash{\SetFigFont{12}{14.4}{rm}y+B}}}
\end{picture}
  \vspace{0.6cm}
  \caption{\label{fig:loop3}A typical graph of $w_\mu^{(3)}(y;AB)$.
  	It starts from $y$, through $y\!+\!A$,
	$y\!+\!B$, and then comes back to $y$.
	The two hypercubic vectors $A$ and $B$ satisfy the condition
	$A_\mu=1$, $B_\mu=0$.}
\end{figure}

\begin{figure}[htbp]
  \setlength{\unitlength}{0.012500in}%
\begingroup\makeatletter\ifx\SetFigFont\undefined
\def\x#1#2#3#4#5#6#7\relax{\def\x{#1#2#3#4#5#6}}%
\expandafter\x\fmtname xxxxxx\relax \def\y{splain}%
\ifx\x\y   
\gdef\SetFigFont#1#2#3{%
  \ifnum #1<17\tiny\else \ifnum #1<20\small\else
  \ifnum #1<24\normalsize\else \ifnum #1<29\large\else
  \ifnum #1<34\Large\else \ifnum #1<41\LARGE\else
     \huge\fi\fi\fi\fi\fi\fi
  \csname #3\endcsname}%
\else
\gdef\SetFigFont#1#2#3{\begingroup
  \count@#1\relax \ifnum 25<\count@\count@25\fi
  \def\x{\endgroup\@setsize\SetFigFont{#2pt}}%
  \expandafter\x
    \csname \romannumeral\the\count@ pt\expandafter\endcsname
    \csname @\romannumeral\the\count@ pt\endcsname
  \csname #3\endcsname}%
\fi
\fi\endgroup
\begin{picture}(88,125)(55,660)
\thinlines
\put(140,759){\circle*{6}}
\put( 60,680){\circle*{6}}
\put( 60,760){\circle*{6}}
\thicklines
\put( 60,759){\line( 1, 0){ 80}}
\special{ps: gsave 0 0 0 setrgbcolor}\put( 58,760){\line( 0,-1){ 80}}
\special{ps: grestore}\special{ps: gsave 0 0 0 setrgbcolor}\put( 61,759){\line( 0,-1){ 79}}
\special{ps: grestore}\special{ps: gsave 0 0 0 setrgbcolor}\put( 58,710){\vector( 0, 1){ 20}}
\special{ps: grestore}\special{ps: gsave 0 0 0 setrgbcolor}\put(110,732){\vector(-1,-1){ 15}}
\special{ps: grestore}\special{ps: gsave 0 0 0 setrgbcolor}\put( 61,710){\vector( 0, 1){ 20}}
\special{ps: grestore}\special{ps: gsave 0 0 0 setrgbcolor}\put(112,730){\vector(-1,-1){ 15}}
\special{ps: grestore}\special{ps: gsave 0 0 0 setrgbcolor}\put( 61,679){\line( 1, 1){ 80}}
\special{ps: grestore}\special{ps: gsave 0 0 0 setrgbcolor}\put( 58,680){\line( 1, 1){ 81}}
\special{ps: grestore}\put( 86,759){\vector( 1, 0){ 20}}
\put( 60,660){\makebox(0,0)[lb]{\smash{\SetFigFont{12}{14.4}{rm}y}}}
\put( 55,770){\makebox(0,0)[lb]{\smash{\SetFigFont{12}{14.4}{rm}y+A}}}
\put(135,770){\makebox(0,0)[lb]{\smash{\SetFigFont{12}{14.4}{rm}y+B}}}
\end{picture}
  \vspace{0.6cm}
  \caption{\label{fig:loop4}A typical graph of $w_\mu^{(4)}(y;AB)$.
  	It starts from $y$, through $y\!+\!A$,
  	$y\!+\!B$, and then comes back to $y$.
	The two hypercubic vectors $A$ and $B$ satisfy the condition
	$A_\mu=0$, $B_\mu=1$.}
\end{figure}

\begin{figure}[htbp]
  \vspace{1.5cm}
  \hspace{1.2cm}
    \epsffile{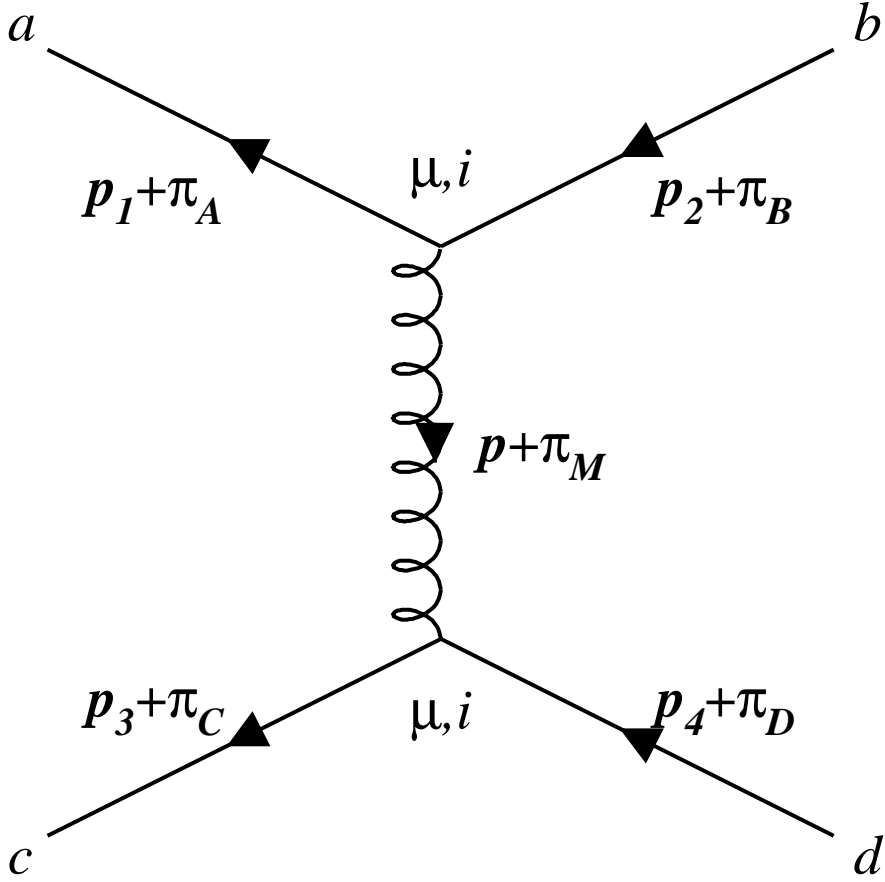}
  \vspace{1.cm}
  \caption{\label{fig:feynman}The Feynman graph which generates
  	four-fermion operators at the tree-level.}
\end{figure}

\end{document}